\documentclass[11pt, oneside]{article}   	% use "amsart" instead of "article" for AMSLaTeX format
\def\be{\begin{eqnarray}}
\def\ee{\end{eqnarray}}

\def\roughly#1{\mathrel{\raise.3ex\hbox{$#1$\kern-.75em%
\lower1ex\hbox{$\sim$}}}}
\def\lsim{\roughly<}
\def\gsim{\roughly>}

\def\bi{\bibitem}

\usepackage{geometry}                		% See geometry.pdf to learn the layout options. There are lots.
\usepackage{graphicx}				% Use pdf, png, jpg, or eps§ with pdflatex; use eps in DVI mode
\usepackage[usenames]{color}
\usepackage{amssymb}
\begin{document}
\hfill {\today }
\vskip 1cm
\centerline{\large\bf  Nuclear Symmetry Energy with Strangeness in Heavy Ion Collision }
\vskip 1cm
\vspace{.30cm}
\begin{center}
Hyun Kyu Lee$^{*}$ and Mannque Rho$^{*,\dagger}$

\vskip 0.50cm
\noindent { \it $^*$Department of Physics, Hanyang University, Seoul 133-791, Korea}

\noindent { \it $^{\dagger}$Institut de Physique Th$\acute{e}$orique, CEA Saclay, 91191 Gif-sur-Yvette, France}

\end{center}
\vskip 1cm
\centerline{\bf ABSTRACT}
\vskip 1cm
The role of anti-kaons in the symmetry energy to be determined in heavy-ion collisions  as for instance in such observables as the $\pi^-/\pi^+$ ratio is discussed using a simple chiral Lagrangian. It is shown, with some mild assumptions, that kaons, when present in the system, can affect the EoS  appreciably for both symmetric and asymmetric nuclear matter. For nuclear matter with small asymmetry with which heavy-ion collisions are studied, it may be difficult to distinguish a stiff symmetry energy and the supersoft symmetry energy, even with kaons present. However the effect of kaon is found to be significant such that $\mu_n-\mu_p \neq 0$ near $x=1/2$, at which the chemical potential difference  is zero without  kaon amplitude.  We present the argument that in order to obtain a reliably accurate equation of state (EoS) for compact-star matter, a much deeper understanding is needed on how the strangeness degrees of freedom such as kaons, hyperons etc. behave in baryonic matter in a Fermi liquid (or possibly a non-Fermi liquid)  phase with potential phase changes. It is suggested that such an {\em accurate} treatment could have an important implication on possibly modified gravity.
\vskip 0.5cm

\newpage

\section{Introduction}
\indent\indent  The new degrees of freedom other than nucleon, such as kaons, hyperons, strongly-coupled quarks etc., in the dense baryonic matter are expected in the heavy ion collision and at the core of  compact stars. The nuclear symmetry energy denoted $E_{sym}$ in the literature figures importantly in nuclear physics and in compact-star physics. The nuclear symmetry energy plays an important role in determining  the  composition of dense baryonic matter  and  controlling the fate of compacts stars, which is one of the principal themes of our current theoretical research into baryonic matter at high density.     Although it is more or less controlled by experiments up to near nuclear matter density $n_0$, it is almost completely unknown at high density relevant to the interior of compact stars going up to, e.g., $\sim 10n_0$~\cite{2Solar}. While there are a large number of theoretical predictions for $E_{sym}$ that range widely above $n_0$, there has been practically no attention paid to the effect of new degrees of freedom, in particular, strangeness, on the nuclear symmetry energy.   For a system of $n$ nucleon number density, the energy differences of the states with different composition of protons and neutrons are encoded in  what we shall call `asymmetry energy', ${\cal E}_{asym}$, defined by subtracting the energy of the state with symmetric compositions, $n_p = n_n= n/2$, from the energy of the system composed of $n_p$  proton number density  and  $n_n$ neutron number density,
\be
{\cal E}_{asym}(n,x)\equiv E(n,x) - E(n,x=1/2)
\ee
with $x = n_p/n$. Empirically it is found that it obeys a parabolic law in the asymmetry factor $ \delta=(1-2x)$ as given by
\be
{\cal E}_{asym}(n,x) = S(n)\delta^2,
\ee
where $S(n)$ is  what is referred to as ``symmetry energy" and conventionally denoted in the literature as $E_{sym}(n)$.\footnote{We use the notation $S(n)$ for the symmetry energy to avoid the confusion with the ``asymmetry energy" denoted ${\cal E}_{asym}$.}

In the presence of strange hadrons,  the asymmetry energy should depend on the strangeness content in the baryonic matter, since there should be nontrivial interactions between nucleon and strange hadrons.  Denoting the strangeness number fraction relative to the nucleon number density as $x_S = n_S/n$, the  asymmetry energy can be modified as
\be
{\cal E}_{asym}(n,x,x_S) \equiv E(n,x,x_S) - E(n,x=1/2,x_S).
\ee
One of the immediate  questions is whether the empirical parabolic law is valid with the presence of the strange degree of freedom, which is one of the motivations of this work.

In this work, in contrast to the interior of compact stars, $x_S$ will be considered as a ``probe parameter" similarly to $x$ and $n$.  While inside the compact star one can assume the weak equilibrium, which would enable us  to determine the strangeness content,  in heavy-ion collisions the transient time for the dense matter phase is too short to activate the weak interactions. This difference between the hadronic matter in heavy ion collisions and the hadronic matter in weak equilibrium of stellar matter renders the role of the symmetry energy different from each other  at some high density. However heavy-ion collisions are expected to probe the same symmetry energy as in compact stars below the density at which kaons start condensing. Here the strange hadrons are produced via strong interactions.  It is, therefore, natural that we cannot expect to have nonzero net strangeness number in an isolated hadronic system, since any process involving net strangeness number production is suppressed.  Nevertheless, there may be possibilities to form a lump of dense baryonic matter with strangeness with hadrons with compensating strangeness escaping from the lump such that the total strangeness produced is zero.  One of the possible scenario is the production of $K^+$ and $K^-$.  Suppose  the $K^-$ is captured in a bound state in a nuclear matter lump due to $K^-N$ attractive interactions, but the $K^+$ escapes out of the lump carrying kinetic energies, thereby cooling  the remaining baryon lump and forming a baryonic lump with finite strangeness number.  It is then expected that the nuclear symmetry energy  of the system (lump) will be modified because of the KN interactions~\cite{LR1}.  This process may or may not occur under the conditions provided by nature -- some of the caveats are given below -- but it illustrates in a clear way how the presence of strangeness can modify the nuclear asymmetry energy, even in neutron-rich systems.

In heavy ion collisions,  the initial  neutron-proton asymmetry factor, $\delta (= 1-2x)$ -- which ranges from $0$ for $^{12}$C  to $0.198$ for $^{197}$Au or 0.227 for $^{238}U$ collisions~\cite{CPM} -- does not change. It is very likely that the baryonic lump with bound kaons has a similar $n$-$p$ asymmetry when it is formed.  Hence  the nuclear symmetry energy around  $x=1/2$ is particularly  of interest and relevance to heavy ion collisions.
One may expect the extra energy to be carried along to cool down the system.  We assume that the s-wave bound state of $K^-$ (that we will simply refer to as ``kaon" unless otherwise noted) is feasible with the amplitude $K$ and energy $E_K$ similarly to kaon condensation\footnote{Here we assume the uniform amplitude inside the lump but vanishing outside, and  we neglect the surface effect.}:
\be
K^- = K \exp^{iE_Kt}.
\ee
For nucleon-nucleon and   kaon-nucleon interactions, we consider a simple -- toy -- model given by the following Lagrangian~\cite{LR1},
\be
{\cal L} = {\cal L}_{KN} + {\cal L}_{NN} \label{Leff}
\ee
where
\be
{\cal L}_{KN} &=& \partial_{\mu} K^- \partial^{\mu} K^+ - m^2_K K^+ K^-  + \frac{1}{f^2} \Sigma_{KN}(\psi_n^{\dagger} \psi_n + \psi_p^{\dagger}\psi_p) K^+ K^-\nonumber \\
&+& \frac{i}{4f^2} (\psi_n^{\dagger} \psi_n + 2 \psi_p^{\dagger}\psi_p) (K^+\partial_{0} K^- - K^- \partial_{0} K^+) +\cdots\label{LKN}\\
{\cal L}_{NN} &=&
   \psi_n^{\dagger} i \partial_{0} \psi_n +  \psi_p^{\dagger} i \partial_{0} \psi_p - \frac{1}{2m_N}( \vec{\nabla} \psi_n^{\dagger}\cdot \vec{\nabla} \psi_n + \vec{\nabla} \psi_p^{\dagger} \cdot \vec{\nabla} \psi_p) - V_{NN}+\cdots\, \label{LNN}
\ee
where the ellipses stand for higher derivatives, higher number of local fields
and the interaction with higher excitations like $\Delta$, hyperon degrees of freedom etc.  $m_N$ ($m_K$) denotes the nucleon (kaon) mass,  $f$ is a constant related to the pion decay constant $f_\pi$, $\Sigma_{KN}$ is the KN sigma term encoding the explicit breaking of chiral symmetry and $V_{NN}$ is an NN potential that we need not specify for our purpose.

The Hamiltonian can be obtained
\be
{\cal H} = {\cal H}_{KN}  + {\cal H}_{NN},
\ee
where
\be
{\cal H}_{KN} &=&  [E_K^2 + m^2_K  -
\frac{n}{f^2} \Sigma_{KN}] K^+ K^-, \label{hkn}\\
{\cal H}_{NN} &=& \frac{3}{5}E_F^0 \left(\frac{n}{n_0}\right)^{2/3}
n + V(n) + n (1-2x)^2 S(n),
\ee
The kaon number density (equivalently strange number density, $-n_S$) is given by
 \be
n_K = \left[2 E_K + \frac{(1+x)n}{2 f^2}\right] K^2 \label{knumber}.
\ee
The kaon chemical potential is determined by the KN interaction given by
 \be
m_K^2 - E_K^2 - E_K \frac{(1+x)n}{2f^2} - \frac{n}{f^2}\Sigma_{KN} = 0. \label{eom}
\ee
Then we get
\be
E_K &=& \left[-\frac{(1+x)n}{2f^2} + \sqrt{(\frac{(1+x)n}{2f^2})^2 - 4(\frac{\Sigma_{KN}}{f^2}n - m_K^2)} \right]/2 \,.\label{mukk},
\ee

\begin{figure}[t!]
\begin{center}
\includegraphics[height=7.0cm]{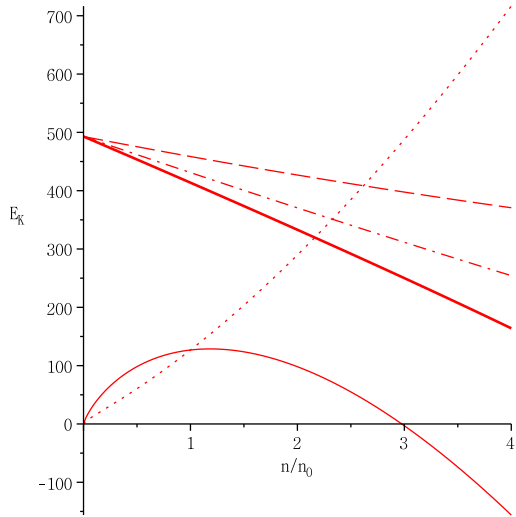}
\vskip 1.5cm
\caption{ The density dependence of $E_K$ with  $\Sigma_{KN}=0, ~~2f$ and $3.2f$ denoted by dashed, dot-dashed and thick solid lines respectively for  $x=0$.  $4S(n)$ (dotted line) and $4S_{ss}(n)$ (thin solid line), roughly the corresponding electron chemical potentials, are plotted together to see the kaon condensation threshold densities in weak equilibrium. With $4S_{ss}(n)$ there is no kaon condensation.}
\label{muksn2}
\end{center}
\end{figure}

The density dependence for given $x$  of $E_K$   is determined by the $KN$ sigma term $\Sigma_{KN}$  and the pion decay constant $f\approx 93$ MeV.  It is given in Fig.~\ref{muksn2} for $\Sigma_{KN} =0, ~~ 2f$ and $3.2 f$ in dashed, dot-dashed and thick solid lines\footnote{The current lattice calculations put a bound on $\Sigma_{KN}\lsim 250$ MeV~\cite{lattice}.} respectively for the pure neutron matter $x=0$.

 Note the steeper slope for larger value  of  $\Sigma_{KN}$.  The  extreme case is for $\Sigma_{KN}=0$, where only the Weinberg-Tomozawa term, the last term in Eq.~(\ref{LKN}), is effective.\footnote{For kaon condensation in star matter for this case, we will need a relatively stiffer symmetry energy.}   Note that the $x$-dependence of $E_K$ is strongly controlled by the Weinberg-Tomozawa term for densities far below the critical condensation density, $n_{th} = m_K^2f^2/ \Sigma_{KN}$,  which comes out to be $\sim 6 n_0$  if one uses  the lattice value for the sigma term.\footnote{  This simple formula based on Eq.~(\ref{Leff}) for $n_{th}$ is perhaps too naive even near -- not to mention far away from -- the equilibrium density of nuclear matter. We are assuming that kaons condense from the state of matter that can be described as a Fermi liquid. To do better, one should approach it with hidden local symmetry Lagrangian in the mean field as mentioned in Section \ref{conclusion}. Such an approach -- in a highly oversimplified form -- was discussed in \cite{BRwalecka} where $n_{th}$ was found to be $\sim 3n_0$.}
This is because the first term in the square-root in Eq.(\ref{mukk}) is much smaller than the second term, $ -4(\frac{\Sigma_{KN}}{f^2}n - m_K^2)$ (which is non-negative).   For  $\Sigma_{KN}$  considered in our model,  the densities to be probed by heavy-ion machines,   $n= 2\sim 3n_0$,  are far from  $n_{th}$.

\section{Results}

The Hamiltonian density in the previous section is simplified to
\be
{\cal H}_{KN} = E_K ~ n_K. \label{hknr}
\ee

The chemical potentials of neutron and proton with bound  s-wave kaon are  given by
\be
\mu_n &=& \mu_n^0 - \left[ \frac{E_K}{2 f^2} + \frac{\Sigma_{KN}}{f^2} \right] K^2, \\
\mu_p &=& \mu_p^0 - \left[ \frac{E_K}{f^2} + \frac{\Sigma_{KN}}{f^2} \right] K^2,
\ee
where
\be
\mu_n^0 - \mu_p^0 = 4 (1 - 2\frac{n_p}{n}) S_N(n).
\ee
 One can notice that the effect of kaon on the neutron and proton chemical potentials
has two parts, one from the Weinberg-Tomozawa term and the other from  $\Sigma_{KN}$ term.  The latter contributes equally to the chemical potentials since it does not depend on the $n-p$ asymmetry\footnote{For the density we consider, we ignore the difference  between $\Sigma_{Kn}$ and $\Sigma_{Kp}$.} while
the former depends on the $n-p$ asymmetry and induces different contributions.  Then we see that  the chemical potential difference $\mu_n^0 - \mu_p^0$ gets an additional contribution from the s-wave kaon amplitude $K$ through Weinberg-Tomozawa term as given by
\be
\mu_n-\mu_p = 4 (1 - 2x) S_N(n) + \frac{E_K}{2 f^2} K^2.\label{Kmunp}
\ee
It should be noted that it is this quantity that one hopes to determine in heavy ion collisions  (at high enough energy to produce kaon pairs), as for example in the ratio of $\pi^+/\pi^-$ and in the future, in $K^+/K^0$ ratio. (See ~\cite{bal}). Given this quantity, then the asymmetry energy per nucleon with bound s-wave kaons can be obtained in the following form,
\be
{\cal E}_{asym}(n,x,x_K) = (1-2x)^2  S_N(n)+[E_K(n,x) -E_K(n,1/2)] x_K, \label{ksym}
\ee
where $x_K= n_K/n$ is the kaon number fraction.

\begin{figure}[t!]
\begin{center}
\includegraphics[height=7.0cm]{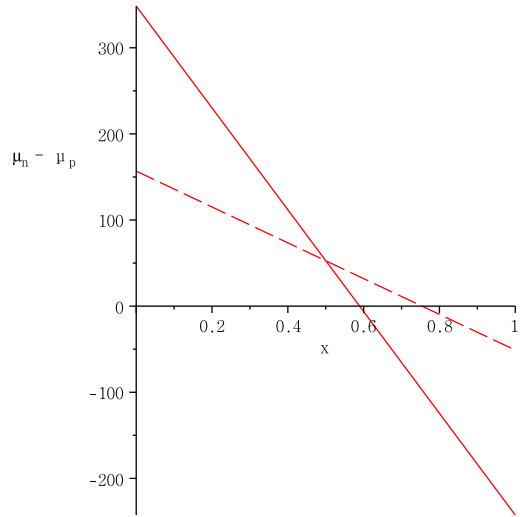}
\vskip 1.5cm
\caption{ The chemical potential difference $\mu_n - \mu_p$ vs. $x$ for $S(n)$ (solid line) and $S_{ss}(n)$ (dotted line). Here, for illustration,  we have taken $n=2n_0$ and the extreme case of $x_K=1$.}
\label{kmunp}
\end{center}
\end{figure}

As a rough estimate of what is involved,  we assume that $S_N(n)$ does not get a substantial modification due to the presence of kaons\footnote{ There is a caveat to this in the case of kaon condensation. As mentioned in Section \ref{conclusion}, condensed  kaon-- perhaps relevant in compact stars, though not in heavy-ion collisions -- could modify significantly the baryon sector.}. Then we may make use of the energy density and symmetry energy factor $S$ of phenomenological models. We take one simple approach called ``MID" (short for ``momentum-independent interaction") used by Li {\it et al.}~\cite{LCK}.\footnote{ The objective of the heavy-ion experiments is to probe the symmetry energy at the range of densities $n\gsim (2-3) n_0$. At such densities and beyond, the detailed structure of interactions, i,.e., tensor forces, short-range repulsions etc., the degrees of freedom involved, e.g., half-skyrmions, kaons, hyperons, strongly-coupled quarks etc. is crucial for understanding the physics of dense compact-star matter. The simple model form we are taking here, though fit to experiments up to $n_0$, is just a parametrization and contains no information of what takes place in the highly correlated matter involved at high density. We use it just to gain a rough idea of what might be going on in two extreme cases.}    We can rewrite the symmetry energy, $S(n)$, of \cite{LCK} as
\be
S(n) &=& (2^{2/3} -1)\frac{3}{5}E_F^0(\frac{n}{n_0})^{2/3} + F \frac{n}{n_0} + (18.6-F)(\frac{n}{n_0})^{C}. \label{2}
\ee
The parameters $A$, $B$ and $\gamma$ determined by experiments are $A= -298.3$ MeV, $B= 110.9$ MeV, $\gamma = 1.21$ and $E_F^0 = 37.2$MeV and  we take  $F=3.673$ and $C=1.569$.
For comparison we consider also a super-soft symmetry energy given by
\be
S_{ss}(n) = 13.1 (\frac{n}{n_0})^{2/3} +107 \frac{n}{n_0} -88.4(\frac{n}{n_0})^{1.25}. \label{ss}
\ee
In Fig.~\ref{kmunp}, the $x$-dependence of the chemical potential difference, Eq.~(\ref{kmunp}),  at $n=2 n_0$ are shown for two symmetry energy factors. The results are very similar for $n=3n_0$. The range of densities $(2-3)n_0$ is appropriate for the purpose because it comes before kaons could condense, and also the baryonic matter could be considered to be in a Fermi liquid state~\footnote{ This is to avoid the possible half-skyrmion phase predicted in the skyrmion crystal model, since the heavy-ion measurements for the meson ratio are not to probe the regime in which topology change could intervene.}.

One can see that the effect of kaon is significant near $x=1/2$, at which the chemical potential difference, the first term in Eq.~(\ref{Kmunp}), is zero without  kaon amplitude. Thus it will   affect the ratios $\pi^-/\pi^0$, $K^+/K_0$ etc.   In the above calculation,  we took $x_K=1$ for simplicity, which is of course too big.  To be realistic, we need to fix it dynamically with the collision condition using a more sophisticated theory as will be mentioned below.  The simplest possibility is the charge neutral baryon lump, which would give $x_K \sim x$  near $ x \sim 1/2$.

The asymmetry energy, ${\cal E}_{sym}$, at $n=2n_0$ for $x_K=1$ is shown  for illustrative purpose for two symmetry energy factors  $S(n)$ and $S_{ss}(n)$ in Fig.~\ref{ksym}.

\begin{figure}[t!]
\begin{center}
\includegraphics[height=7.0cm]{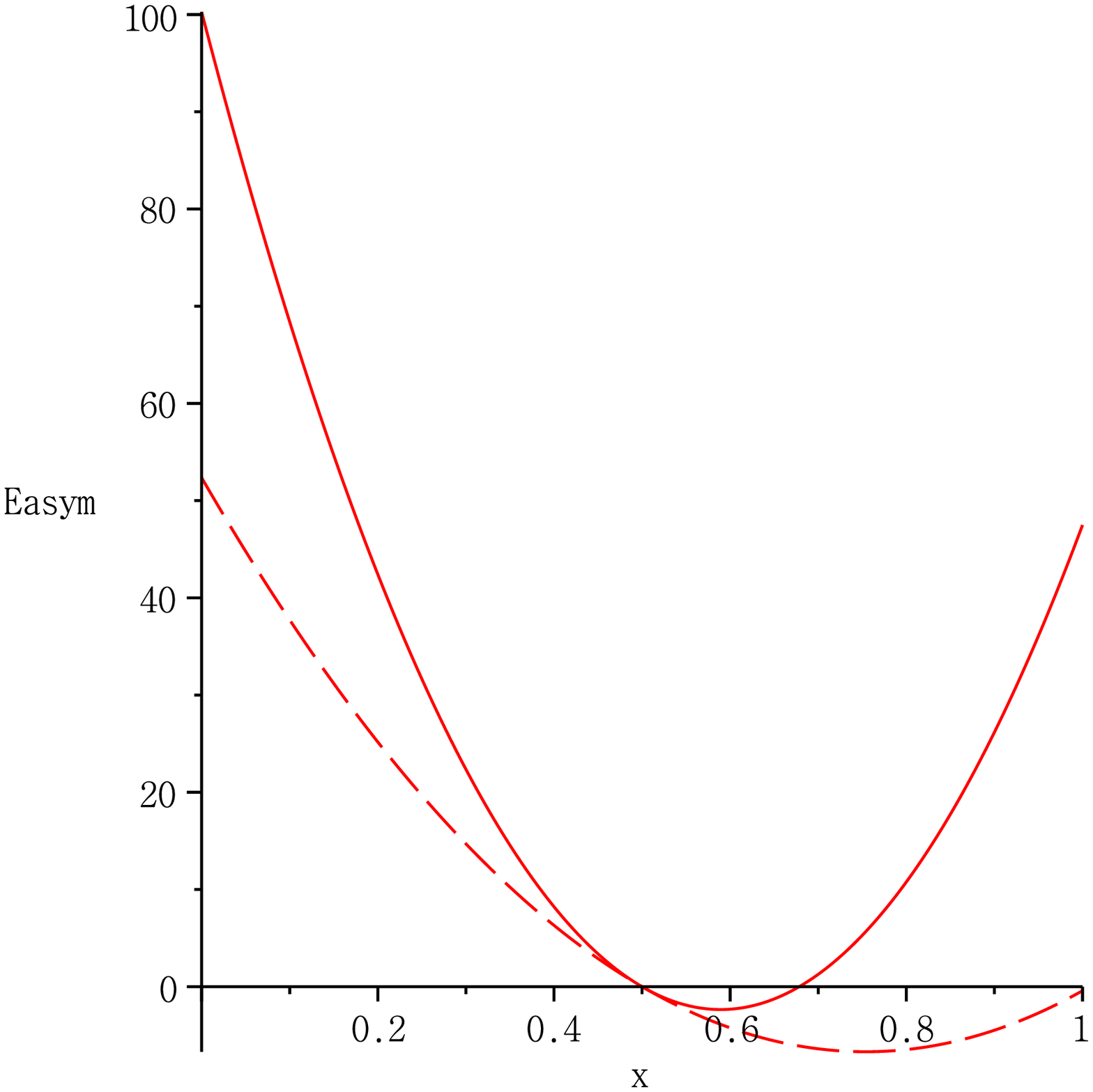}
\vskip 1.5cm
\caption{The asymmetry energy  per nucleon, ${\cal E}_{asym}(n,x, x_K=1)$,  vs. $x$ for $S(n)$ (solid line) and $S_{ss}$ (dotted line) for $n=2n_0$ and $x_K=1$. }
\label{ksym}
\end{center}
\end{figure}

We see that the neutron-proton permutation symmetry in the nucleon sector, characterized by $(1-2x)^2$,  is significantly  distorted in the presence of kaons. The effect is more prominent in the super-soft case. {It is  interesting to note that  the minima, $x_{min}$,  in Fig. \ref{ksym} are  shifted toward $x_{min} > 1/2$, which is equivalent to the proton rich configuration.\footnote{In realistic case,  the minimum of  ${\cal E}_{asym}$ is expected to be further modified when electromagnetic interaction is included.}}
So far the weak equilibrium condition applicable in compact stars has not been used, so the equilibrium threshold density for kaon has no meaning.   Kaons are produced by strong interactions. This is the reason why we can consider kaon amplitude even at lower density $n\lsim 3n_0$, below the weak-equilibrium kaon threshold.

\section{Compact star matter and kaon condensation}\label{conclusion}
In this section we make a few remarks concerning the reliability and relevance of what's discussed above to the principal object of {recent interest}, compact stars.

 The model Lagrangian adopted here, (\ref{LKN}), is limited to the chiral order ${\cal O}(p^2)$ of a chiral Lagrangian that arises from a flavor $SU(3)$ chiral Lagrangian that contains, in addition to the octet baryons and pseudo-Goldstone bosons, $U(3)$ vector mesons, when the hyperons and the vector mesons are integrated out and only the leading terms (both in derivatives and number of fields) are retained. To that order, the kaon number density is quadratic in the kaon field amplitude.  Now one may wonder what happens if higher kaon fields figure in the Lagrangian as is expected in the derivative expansion in the integrating-out . They would enter in $n_K$. The question can be raised as to whether kaon interactions mediated through coupling with the nucleons would not generate repulsion that would limit to and saturate at a finite $n_K$~\cite{gal}.

This question can be addressed with a simple RG argument with (\ref{LKN}) for heavy-ion systems, where the effective kaon mass cannot go down so much. In terms of chiral perturbation theory, the leading order term in the $KN$ interactions is the Weinberg-Tomozawa term as used in the literature~\cite{gal}. This term is ``irrelevant" in the RG flow. Thus, at least in the perturbative sense -- which should be reliable for density not so high above $n_0$, higher kaon field operators cannot do much.

 For a compact star matter, in contrast to the heavy ion collision process, the weak interaction  becomes sufficiently active  to drive  the star matter to be stabilized in weak equilibrium.  In weak equilibrium, the chemical potential of kaon, $\mu_K$, should be identical to electron chemical potential, $\mu_e$:
\be
\mu_e = \mu_K = \mu_n - \mu_p  \label{munpk}.
\ee
When $\mu_K(=\mu_e)$  crosses   $E_K$ in Fig. 1, kaon condensation can occur and it defines the threshold density for kaon condensation, $n_{th}$.
The strange number fraction after kaon condensation threshold should range from zero (condensation threshold) to 1/2 (masseless kaon) for a locally neutral star matter in weak equilibrium.
Since the kaon number fraction is not larger than 1/2, the attractive nature of $KN$ interactions may remain  dominant over  the repulsion between kaons such that it  might not be  strong enough  to destroy the $KN$ attraction for the kaon condensation\footnote{  In the molecules or solid, most of electrons are bound to their parent nuclei, although the interactions between  electrons inside an atom or in different atoms are repulsive electromagnetically.  Naively, it may be that  the shared kaons even mediate more  attractions between nucleons.}.  This could also be the case in  heavy-ion collisions for a formation of kaon bound baryon lump~\cite{gal}.

It is possible that the dynamics involved could very well be inaccessible by the mean-field approach even at not too high density as suggested in \cite{yamazaki}. For instance, the smooth transition at about $\sim 2n_0$ from hadronic matter to strongly interacting quark matter in compact star matter that can accommodate the $\sim 2 M_\odot$ stars~\cite{hatsuda} resembles the kaon condensation scenario associated with the topology change that takes place at $\sim 2n_0$  as suggested in \cite{half-skyrmions,MR-GEB} and mentioned below. Furthermore there is a strong indication that kaon condensation and hyperon appearance should be considered on the same footing~\cite{LR-hyperon}. In the literature, the two processes have been treated separately with no connection between the two. This point may also be pertinent for the process in heavy ion collisions discussed here.

Practically all the treatments of kaon condensation in compact stars available in the literature are anchored on chiral Lagrangians of the type given in (\ref{Leff}). With the Lagrangian of the form (\ref{Leff}) treated in the mean field, implemented naively into dense baryonic systems, the kaons are supposed to start to condense at the density $n_{th}$ where $E_K$ crosses the electron chemical potential in weak equilibrium as shown in Figure \ref{muksn2}. Interpreted in this way, it is generally believed that the EoS of compact star matter is too soft to support the star accurately measured in \cite{2Solar}. In fact with the parameters of the Lagrangian compatible with what's given by experiments, treated in the mean field, as suggested by Brown and Bethe~\cite{brown-bethe}, a star of mass greater than 1.5$M_\odot$ would tend to collapse to a black hole.

The recent accurate measurements of massive $\gsim 2$ solar-mass stars~\cite{2Solar} seem to go against the Brown-Bethe scenario and are in fact taken in some circles to ``rule out" kaon condensation in compact stars. We will argue that this conclusion is unfounded and premature. This is because given the extremely poor knowledge we have on the state of baryonic matter at high density,  the presently available theoretical tools to treat strangeness in dense compact star matter are far from trustworthy.

There are several reasons for our claim that there is a dire need for more unified and consistent treatment of the problem. We cite a few examples. One is that when chiral Lagrangians are put on crystal lattice to simulate dense matter, the only theoretical tool available at present that has a contact with QCD, albeit for large $N_c$, one finds that there can be a topology change from a baryonic matter in the form of skyrmions to one of half-skyrmions~\cite{half-skyrmions}. Such a phase change is not present in the currently employed approaches, i.e., those based on chiral perturbation theory.  It is nonpertubative. In its presence,  the usual mean-field approximation used both for  kaon condensation and for hyperons, two principal sources for strangeness, could very well break down~\cite{MR-GEB}.
 If this is the case, then the question will be how to formulate kaon condensation in  a matter where the mean-field treatment of the baryonic matter cannot be trusted.

 Another possible mechanism to consider is that when kaons condense as in the case of compact-star matter where the electron chemical potential enters in triggering kaon condensation~\cite{novel}, it may be necessary to consider carefully how the condensing kaons couple to the Fermi liquid or possibly to a non-Fermi-liquid baryonic matter~\cite{PR2014}. For instance, in condensed matter, ``critical boson"\footnote{ Here ``critical boson" in the condensed matter terminology means the boson whose mass is tuned to vanish.}-Fermi liquid fermion coupling can lead to a phase change to a non-Fermi-liquid state with various thermodynamic properties severely altered~\cite{kachru}.  Normally, Goldstone-boson coupling to Fermi liquid is irrelevant in the RG sense, since the coupling is of derivative type. In that case, such a breakdown is not expected to occur even though the Goldstone bosons are gapless. Now the kaon is a pseudo-Goldstone boson, so could the renormalization group property of kaon condensation be affected by its irrelevant derivative coupling? The answer to this question is not obvious.  First of all, kaons condense because the chiral symmetry is broken. In the naive interpretation of the chiral Lagrangian (\ref{Leff}), i.e., at the tree level, the KN sigma term determines the threshold density: the larger the KN sigma term -- and hence the greater symmetry breaking, the lower the critical condensation density. Furthermore, in the successful skyrmion description of hyperons as kaon-soliton bound states (without~\cite{klebanov} or with~\cite{scoccola} hidden gauge mesons), the kaon is taken to be massive. In this description, it is the Wess-Zumino term that dominates in binding the kaon to the soliton, with the Wess-Zumino term playing a role equivalent to the Weinberg-Tomozawa term, which is of the leading order in the chiral counting in effectivel Lagrangians. What is also quite intriguing is that by dialing the kaon mass to zero, one can go from the bound-state description to the $SU(3)$ skyrmion description. Now in the presence of vector mesons in hidden local symmetric theory with the vector manifestation, the sigma term could even become relevant.  This would produce instability in the Fermi-liquid structure of the baryonic matter of the type discussed above.

 Suppose that a future theory of kaon condensation formulated {\em reliably and fully consistently} in both mesonic and baryonic sectors gives an EoS that is still too soft to accommodate the observed massive stars with the standard theory of gravity. Then one could envisage a  possibility that a change in fundamental physics is required, with the EoS of compact stars providing a serious indication for a modified gravity. In fact, a suitably modified gravity such as for instance $f(R)$ gravity is argued to be able to  accommodate the massive stars~\cite{odintsov}.

\section{Discussion}

 In this work, we have investigated the effect of the strangeness degree of freedom on the symmetry energy assuming  baryon lumps with kaons bound, which might be produced in heavy ion collision.  One of the possible scenario is that the $K^-$ is captured in a bound state in a nuclear matter lump due to $K^-N$ attractive interactions, but the $K^+$ escapes out of the lump carrying kinetic energies, thereby forming a baryonic lump with finite strangeness number.  We take the simplest  Lagrangian to describe this system, Eq.(\ref{Leff}). It is then expected that the nuclear symmetry energy  of the system (lump) will be modified because of the KN interactions.  It is found
 that even in the presence of kaons, there is little difference in  the asymmetry energies with $S$ and $S_{ss}$ near $x \sim 1/2$, which is roughly the initial condition of heavy-ion collisions. Pertinent to experimental efforts, it should be kept in mind that if our result is correct, the pion ratio could not distinguish between a stiff symmetry energy and a super-soft symmetry energy. The presence of kaons, however, distorts the fermi levels of neutrons and protons via the Weinberg-Tomozawa term such that $\mu_n-\mu_p \neq 0$ at $x=1/2$, which  should vanish without kaons.  It will affect the particle spectrum including the pion ratio in the heavy ion collision.

There are a few caveats.  The scenario in this work is  that $K^-$ is bound to the nuclear matter, but $K^+$ escapes out of  the nuclear matter.  Of course, the crucial question is then how to control the kaon number, a problem yet to be worked out.  Since we take the simplest form of Lagrangian in this work,  the possible roles of higher derivatives,     higher dimension field operators  and moreover the interaction with higher excitations such as $\Delta$,  $\Lambda (1405)$, hyperons etc. should be discussed in detail, which remains as a future work.

The relevant topic to the above discussion is the possibility and the effect of  kaon condensation on the compact star matter.   One of the possibilities is that kaon condensation would produce instability in the Fermi-liquid structure of the baryonic matter. This phenomenon may be relevant perhaps only very near the critical condensation density $n_{th}$, but it indicates the possibility of a variety of subtleties in the EoS as one goes above a few times the normal density $n_0$. And therefore it would be extremely interesting to experimentally observe precursor phenomena to the possible breakdown of Fermi liquid structure discussed above.

\subsection*{Acknowledgments}
We are grateful for discussions with Bao-An Li and Won-Gi Paeng.
This work was partially supported by the WCU project of Korean Ministry of
Education, Science and Technology (R33-2008-000-10087-0).

\end{document}